\documentclass[graybox]{svmult}

\usepackage{makeidx}         
\usepackage{graphicx}        
\usepackage{multicol}        

\usepackage{latexsym}
\usepackage{amssymb}
\usepackage{amsmath}

\def\R{{\Bbb R}}
\def\C{{\Bbb C}}

\def\diag{{\rm diag}}
\def\cl{{\cal C}\!\ell}
\def\even{{\rm even}}
\def\odd{{\rm odd}}
\def\I{\mathbb{I}}

\def\proof{\medbreak\noindent{\bf Proof}}

\def\E{\mathbb{E}}

\def\T{{\rm T}}

\makeindex             

\begin{document}

\title*{Model Dirac and Dirac-Hestenes equations\\ as covariantly equipped systems of equations}

\author{Marchuk N.G.}
\institute{Nikolay Marchuk \at Steklov Mathematical Institute of Russian Academy of Sciences, 8 Gubkina St. Moscow,
119991, Russia \email{nmarchuk@mi-ras.ru}}

%
%
\maketitle

\begin{abstract}
    -We define a new class of partial differential equations of first order (complex covariantly equipped systems of equations), which are invariant with respect to (pseudo)orthogonal changes of cartesian coordinates of (pseudo)euclidian space. It is shown that for pseudoeuclidian spaces of signature (1,n-1) covariantly equipped systems of equation can be written in the form of Friedrichs symmetric hyperbolic systems of equations of first order. We prove  that Dirac and Dirac-Hestenes model equations belong to the class of covariantly equipped systems of equations.
\end{abstract}

\abstract*{
We define a new class of partial differential equations of first order (complex covariantly equipped systems of equations), which are invariant with respect to (pseudo)orthogonal changes of cartesian coordinates of (pseudo)euclidian space. It is shown that for pseudoeuclidian spaces of signature (1,n-1) covariantly equipped systems of equation can be written in the form of Friedrichs symmetric hyperbolic systems of equations of first order. We prove  that Dirac and Dirac-Hestenes model equations belong to the class of covariantly equipped systems of equations.
}
\medskip

Key words: Dirac equation, differential forms, Clifford algebra, genform, symmetric hyperbolic systems, covariantly equipped
\medskip

MSC Class: 35Q41, 83A05

\vskip1cm

\tableofcontents
\medskip

In \cite{TMPh2019} we define a class of so-called {\em real covariantly equipped systems of equations of first order} and prove that a Cauchy problem for  Maxwell's system of equations is equivalent to a Cauchy problem for some covariantly equipped system of equations.
In what follows we define a class of so-called {\em complex covariantly equipped systems of equations of first order}. We see that the Dirac-Hestenes equation can be considered as a complex covariantly equipped system of equations. Also we prove that a Cauchy problem for  Dirac equation (with a nonabelian gauge symmetry) is equivalent to  a Cauchy problem for some complex covariantly equipped system of equations.


\section{Complex covariantly equipped systems of equations of first order}

\noindent{ \bf (Pseudo)euclidian space $\R^{r,s}$}.
 Let $n$ be a natural number, $r,s$ be nonnegative integer numbers, and $\R^{r,s}$ be the $n=r+s$ dimensional  (pseudo)euclidian space with cartesian coordinates  $x^\mu$, $\mu=1,\ldots,n$ and with a metric tensor given by the diagonal matrix
$\eta=\|\eta_{\mu\nu}\|=\diag(1,\ldots,1,-1,\ldots,-1)$ with $r$ pieces of $1$ and $s$ pieces of $-1$ on the diagonal.
\medskip

\noindent{\bf Genforms}. In \cite{tr_mian2019} we define a set (an algebra) of genforms $\Lambda^{[h]}(\R^{r,s})$ considered as a special geometrized representation of Clifford algebra. We use notations from  \cite{tr_mian2019}. Let $e^1,\ldots,e^n$ be generators of Clifford algebra  $\cl(r,s)$ and $e$ be the identity element. And let $y^\mu_a$ be a tetrad in $\R^{r,s}$. We may take {\em a genvector}
 $h^\mu := y^\mu_a e^a\in \cl_1(r,s)\T^1$. Consider the set of genforms
$$
\Lambda^{[h]}(\R^{r,s})=\Lambda^{[h]}_\even(\R^{r,s})\oplus\Lambda^{[h]}_\odd(\R^{r,s})=
\bigoplus_{k=0}^n\Lambda_k^{[h]}(\R^{r,s}),
$$
where
\begin{eqnarray*}
&& \Lambda^{[h]}_\even(\R^{r,s}) = \Lambda^{[h]}_0(\R^{r,s})\oplus\Lambda^{[h]}_2(\R^{r,s})\oplus\ldots,\\
&& \Lambda^{[h]}_\odd(\R^{r,s}) = \Lambda^{[h]}_1(\R^{r,s})\oplus\Lambda^{[h]}_3(\R^{r,s})\oplus\ldots.
\end{eqnarray*}
We see that a genfrom $U=U(x)\in\Lambda^{[h]}(\R^{r,s})$ of the form
\begin{equation}
U = u e + \sum_{k=1}^n\frac{1}{k!} u_{\mu_1\ldots\mu_k}h^{\mu_1}\wedge\ldots\wedge h^{\mu_k}
= u e + \sum_{k=1}^n\sum_{\mu_1<\ldots<\mu_k} u_{\mu_1\ldots\mu_k}h^{\mu_1\ldots \mu_k}\label{U:genform}
\end{equation}
represents a set of real covariant antisymmetric  tensor fields $u,u_\mu,u_{\mu_1\mu_2}$, $\ldots, u_{\mu_1\ldots\mu_n}$, which are, generally speaking, depend on  $x\in\R^{r,s}$. If we have a set of complex covariant antisymmetric  tensor fields $u,u_\mu,u_{\mu_1\mu_2},\ldots u_{\mu_1\ldots\mu_n}$, then it can be represented by complexified genform (\ref{U:genform}). In this case we write $U\in\C\otimes\Lambda^{[h]}(\R^{r,s})$.

In what follows we use the differential operator $\eth=h^\mu\partial_\mu$, which was discussed  in \cite{tr_mian2019}.
\medskip

\noindent{\bf Complex covariantly equipped systems of equations of first order.}\label{sec:SI}
In (pseudo)euclidian space $\R^{r,s}$ with cartesian coordinates $x^\mu$ we consider  systems of differential equations of first order of the form
\begin{equation}
\eth\phi + Q(\phi) = f,\label{sis}
\end{equation}
which are invariant with respect to changes of coordinates from the group $O(r,s)$. In (\ref{sis})
  $\phi=\phi(x)\in \C\otimes\Lambda^{[h]}(\R^{r,s})$ is an unknown genform, $f=f(x)\in \C\otimes\Lambda^{[h]}(\R^{r,s})$ is a known genform of right hand part. Suppose that the term  $Q(\phi)\in \C\otimes\Lambda^{[h]}(\R^{r,s})$ is of the form
\begin{equation}
Q(\phi) = \sum_{j=1}^N A_j\phi B_j,\label{Q:phi}
\end{equation}
and $A_1,\ldots,A_N$; $B_1,\ldots,B_N$ are some known genfroms, which are independent of $\phi$. In addition to this main case with $\phi,f,Q(\phi)\in\C\otimes\Lambda^{[h]}(\R^{r,s})$ we consider two partial cases of equations (\ref{sis}). Namely
\begin{eqnarray}
&& \phi\in\C\otimes\Lambda^{[h]}_\even(\R^{r,s}),\quad f,Q(\phi)\in\C\otimes\Lambda^{[h]}_\odd(\R^{r,s}),\label{case:even}\\
&& \phi\in\C\otimes\Lambda^{[h]}_\odd(\R^{r,s}),\quad f,Q(\phi)\in\C\otimes\Lambda^{[h]}_\even(\R^{r,s}).\label{case:odd}
\end{eqnarray}
In these partial cases in (\ref{Q:phi}) every term $A_j\phi B_j$ is such that $A_j\in\C\otimes\Lambda^{[h]}_\even(\R^{r,s})$,
$B_j\in\C\otimes\Lambda^{[h]}_\odd(\R^{r,s})$, or  $A_j\in\C\otimes\Lambda^{[h]}_\odd(\R^{r,s})$,
$B_j\in\C\otimes\Lambda^{[h]}_\even(\R^{r,s})$.


\section{Covariantly equipped systems of equations in $\R^{1,n-1}$ as a subset of the set of Friedriechs symmetric hyperbolic systems of equations of first order}

\noindent{\bf Friedriechs symmetric hyperbolic systems of equations of first order.}
Let $n\geq2$ be even integer number and let $\R^n$ be the euclidian space with cartesian coordinates $x^1,\ldots,x^n$. Consider a bounded open domain $\Omega\subset\R^n$  such that $x^1>0$. We are interested in a Cauchy problem for a linear system of differential equations of first order
\begin{eqnarray}
&& \sum^n_{i=1} H_i\partial_i u+ Q u=j,\quad x\in\Omega,\label{symhyp}\\
&& u=\psi,\quad x\in S,\label{init:cond}
\end{eqnarray}
 where $H_1,\ldots,H_n,Q$ are complex square matrices of dimension  $N$, which smoothly depend on  $x=(x^1,\ldots,x^n)\in\R^n$; $u=u(x)$ is a $N$-dimensional complex vector-column of unknown functions; $j=j(x)$ is  a $N$-dimensional complex vector-column of functions of right hand part; $S$ is an open domain of the plane $x^1=0$, which we can get from  $n-1$-dimensional domain $\bar\Omega\cap(x^1=0)$ if we remove boundary (we suppose that this boundary is smooth); $\psi=\psi(x^2,\ldots,x^n)$ is a $N$-dimensional complex vector-column of functions of initial data (on $x^1=0$).

 If for all $x\in\Omega$ matrices $H_1,\ldots,H_n$ are Hermitian ($H_i^\dagger=H_i$) and the matrix $H_1$ is positive defined, then the system of equations of first order (\ref{symhyp}) is called {\em a complex Friediechs symmetric hyperbolic system of equations of first order} in domain $\Omega$. We will use the Hermitian scalar product
$$
(\xi,\eta)=\sum_{j=1}^N\xi_j\bar\eta_j.
$$
Suppose that for the Cauchy problem (\ref{symhyp}), (\ref{init:cond}) the following conditions are satisfied:
\begin{itemize}
\item{a)} Matrices $H_i$ are Hermitian, the matrix $H_1$ is positive defined and there exists a constant $\gamma>0$ such that $(H_1\xi,\xi)>\gamma(\xi,\xi)$ for all nonzero complex $N$-dimensional vectors $\xi$ and for all $x\in\Omega$.
    \item{b)} For $x^1>0$ the closure of domain $\bar\Omega\subset\R^n$ is bounded by the surface $\partial\Omega$. For this surface a vector of exterior normal $\tau=(\tau_1,\ldots,\tau_n)$ at any point $x\in\partial\Omega$ is such that the matrix $\sum_{i=1}^n H_i\tau_i$ is positive defined.
             \item{c)} The functions with values in martices
        $$
        H_i=H_i(x),\quad Q=Q(x),\quad j=j(x),\quad x\in\Omega
        $$
        are smooth (infinitely differentiable) functions of $x\in\Omega$. The boundary $\partial\Omega$ of the domain $\bar\Omega$ for $x^1>0$ is smooth. The vector-function $\psi=\psi(\acute x)$ is a smooth function of $\acute x\in S$. The boundary of the domain $\bar S$ is smooth.
                \end{itemize}

                According to the theory of Friediechs symmetric hyperbolic systems of first order, if conditions a),b),c) are satisfied, then there exists a classical (continuously differetiable) solution of the Cauchy problem (\ref{symhyp}), (\ref{init:cond}). Also, there is an a priory estimate that gives us correctness of the Cauchy problem (see  \cite{Fri},\cite{Godunov1979},\cite{Miz}).
\medskip

Now we want to write down a covariantly equipped system of equations in pseudoeuclidian space $\R^{1,n-1}$ in the form of Friediechs symmetric hyperbolic system of first order. The operation of Hermitian conjugation of genforms in $\R^{1,n-1}$ was considered in details in the book \cite{Mybook2018} (section 3.9). For an arbitrary genform $V\in\C\otimes\Lambda^{[h]}(\R^{1,n-1})$ the Hermitian conjugated genform is defined by the formula
$$
V^\dagger := \beta\tilde{\bar V}\beta,
$$
 where $\beta=e^1$ is the generator of Clifford algebra $\cl(1,n-1)$ such that $(e^1)^2=e$. By $\tilde V$ we denote the  reverse operation and by $\bar V$ we denote the complex conjugation operation. For a genvector $h^\mu$ we have
\begin{equation}
(\beta h^\mu)^\dagger=\beta h^\mu,\quad \mu=1,\ldots,n.\label{beta:h}
\end{equation}
We may multiply both parts of the system of equation (\ref{sis}) from left by the genform $\beta$ and, as a result, we get the system of equations
\begin{equation}
\beta h^\mu\partial_\mu\phi + \beta Q(\phi) = \beta f,\label{beta:sis}
\end{equation}
which is symmetric according to the identity
 (\ref{beta:h}). To prove the hyperbolicity  in Friedrichs sense of the system of equations (\ref{beta:sis}) we must prove that the genform $\beta h^1$ is positive defined. For Minkowski space $\R^{1,3}$  this proposition was proved in \cite{Mybook2018} (section 5.7, Theorem 5.2). It is easy to check that the proof of Theorem 5.2 in \cite{Mybook2018} is valid for an arbitrary even $n\geq2$. Finally, to get a connection with the theory of Friedrichs symmetric hyperbolic systems of first order, we can use a matrix representation of Clifford algebra $\cl(1,n-1)$ that consistent with the operation of Hermitian conjugation of genforms (see \cite{Mybook2018}, section 11.3).

\medskip

So, we see that a complex covariantly equipped system of equations can be written in the form of Friedrichs symmetric hyperbolic system of equations of first order.

\section{Modified Dirac-Hestenes equation as a covariantly equipped system of equations}

In \cite{ILeq}-\cite{ObSol} one consider a modification of the Dirac equation of an electron, which uses differential forms and a wave function has 16 complex components. D.~Hestenes \cite{Hestenes} has found a new form of the Dirac equation with the aid  of even elements of Clifford algebra $\cl(1,3)$. A modification of the Dirac-Hestenes equation using even and odd genforms was considered in \cite{Mybook2018} (section 9.1). This modification in Mikowski space $\R^{1,3}$ looks as follow
\begin{equation}
h^\mu(\partial_\mu\Psi+a_\mu\Psi K)+ m\Psi K\beta=0,\label{MDH1}
\end{equation}
where $\Psi\in\Lambda^{[h]}_\even(\R^{1,3})$ or  $\Psi\in\Lambda^{[h]}_\odd(\R^{1,3})$, $K=-e^{23}$, $\beta=e^1$.
Note that a genform $K$ satisfy conditions
$$
K\in \Lambda^{[h]}_2(\R^{1,3}),\quad K^2=-e,\quad [\beta,K]=0.
$$
So, as a $K$ we may take $\pm e^{23}$, $\pm e^{24}$, $\pm e^{34}$ or a linear combination of these elements with real coefficients (and with the condition $K^2=-e$). Using the operator  $\eth=h^\mu\partial_\mu$ and a genform $A=a_\mu h^\mu\in\Lambda^{[h]}_1(\R^{1,3})$, the equation (\ref{MDH1}) can be written in the form of covariantly equipped system of equations
\begin{equation}
\eth\Psi + A\Psi K + m\Psi K\beta=0.\label{MDH2}
\end{equation}
\medskip

According to results of previous section, the system of equations (\ref{MDH2}) can be written as Friedrichs symmetric hyperbolic system of first order. This leads to the correctness of Cauchy problem for (\ref{MDH2}) in a proper domain of Minkowski space.

\section{How to reduce a Cauchy problem for a model Dirac equations with nonabelian gauge symmetry to a Cauchy problem for covariantly equipped system of equations}

\noindent{\bf Hermitian idempotents and related structures.} An element (scalar) $t\in\C\otimes\Lambda^{[h]}(\R^{r,s})$ that satisfy conditions
\begin{equation}
t^2=t,\quad t^\dagger=t,\quad \partial_\mu t=0,\  (\mu=1,\ldots,n),\label{herm:idem}
\end{equation}
is called {\em a Hermitian idempotent (projector)}.
 \footnote{It was proved in \cite{Mybook2018} (section 3.12) that for Minkowski space $\R^{1,3}$ in the complexified Clifford algebra $\C\otimes\cl(1,3)$ there are five types of Hermitian idempotents
 \begin{eqnarray*}
 && t_{(0)} = 0,\quad t_{(1)} = \frac{1}{4}(e+e^1+i e^{23}+i e^{123}),\quad t_{(2)} = \frac{1}{2}(e+e^1),\\
 && t_{(3)} = \frac{1}{4}(3 e+ e^1+i e^{23}-i e^{123}),\quad t_{(4)} = e.
 \end{eqnarray*}
}

 Let us introduce the following sets of genforms from
$\C\otimes\Lambda^{[h]}(\R^{r,s})$:
\begin{eqnarray*}
I(t) &=& \{U\in\C\otimes\Lambda^{[h]}(\R^{r,s}) : U = U t\},\\
K(t) &=& \{U\in I(t) : U=t U\},\\
L(t) &=& \{U\in K(t) : U^\dagger= - U\},\\
G(t) &=& \{U\in\C\otimes\Lambda^{[h]}(\R^{r,s}) : U^\dagger U=e,\ U-e\in K(t)\}.
\end{eqnarray*}
These sets of genforms $I(t),K(t),L(t),G(t)$ can be interpreted as follows: $I(t)$ is a left ideal; $K(t)$ is two-sided ideal; $L(t)$ is a real Lie algebra of a unitary  Lie group; $G(t)$ is a unitary Lie group.

\medskip

\noindent{\bf A model Dirac equation.} {\em A model Dirac equation} (\cite{Mybook2018}, chapter 5) in a (pseudo-)Euclidian space $\R^{r,s}$ with a fixed Hermitian idempotent $t\in\C\otimes\Lambda^{[h]}(\R^{r,s})$ can be written in the form
\begin{equation}
h^\mu(\partial_\mu\psi + \psi A_\mu) + i m \psi = 0,\label{mod:Dirac:eq}
\end{equation}
 where $h^\mu$ is a genvector, $\psi=\psi(x)\in I(t)$, $A_\mu=A_\mu(x)$ are components of a covector with values in  $L(t)$, $i$ is the imaginary unit, $m$ is a real constant (a mass of a particle). Note that the genform $\psi=\psi(x)\in I(t)$, which is unknown in equation (\ref{mod:Dirac:eq}), is supposed to be a scalar in $\R^{r,s}$.
\medskip

\noindent{\bf A procedure of covariant equipment of a Cauchy problem for the model Dirac equation.} According to the definition (in the first section) of a covariantly equipped system of equations, we see that system of equations (\ref{mod:Dirac:eq}) is a covariantly equipped system of equations iff $t=e$. What to do for Hermitian idempotents $t\neq e$?

In what follows, we restrict ourselves by pseudo-Euclidian spaces  $\R^{1,n-1}$ with even $n\geq2$. Consider a Cauchy problem for system of equations (\ref{mod:Dirac:eq})
\begin{eqnarray}
&& h^\mu(\partial_\mu\psi + \psi A_\mu) + i m \psi = 0,\quad x^1>0\label{mod:Dirac:eq1}\\
&& \psi = \psi_0,\quad x^1=0,\label{init:cond1}
\end{eqnarray}
 where $\psi_0=\psi_0(x^2,\ldots,x^n)=\psi_0 t\in I(t)$ are a known finite smooth genform of initial data of the Cauchy problem. With Cauchy problem (\ref{mod:Dirac:eq1}), (\ref{init:cond1}) one can associate the following Cauchy problem for a covariantly equipped system of equations:
\begin{eqnarray}
&& h^\mu(\partial_\mu\Psi + \Psi A_\mu) + i m \Psi = 0,\quad x^1>0\label{mod:Dirac:eq2}\\
&& \Psi = \psi_0,\quad x^1=0,\label{init:cond2}
\end{eqnarray}
where $\Psi=\Psi(x)\in\C\otimes\Lambda^{[h]}(\R^{1,n-1})$, $x\in\R^{1,n-1}$.

\begin{theorem}
\begin{itemize}
\item{I)} If a genform $\psi \in I(t)$ is a solution of Cauchy problem (\ref{mod:Dirac:eq1}),(\ref{init:cond1}), then the genform $\Psi\equiv\psi\in \C\otimes\Lambda^{[h]}(\R^{1,n-1})$ satisfies Cauchy problem (\ref{mod:Dirac:eq2}),(\ref{init:cond2}).
\item{II)} If a genform  $\Psi\in \C\otimes\Lambda^{[h]}(\R^{1,n-1})$ is a solution of Cauchy problem (\ref{mod:Dirac:eq2}),(\ref{init:cond2}), then the genform $\Psi$ belongs to the set $I(t)$ and the genform $\psi=\Psi$ satisfies Cauchy problem  (\ref{mod:Dirac:eq1}),(\ref{init:cond1}).
\end{itemize}
\end{theorem}
\proof. The proposition I) of theorem is evident. Let us prove the proposition II). For a Hermitian idempotent
 $t\in\C\otimes\Lambda^{[h]}(\R^{1,n-1})$ one can define {\em the dual Hermitian idempotent}
$$
t^\prime := e - t,
$$
 where $e$ is the identity element of Clifford algebra $\cl(1,n-1)$ and simultaneously $e$ is the identity element of the set of genforms $\C\otimes\Lambda^{[h]}(\R^{1,n-1})$. Let a genform $\Psi\in \C\otimes\Lambda^{[h]}(\R^{1,n-1})$ be a solution of Cauchy problem (\ref{mod:Dirac:eq2}),(\ref{init:cond2}).
 One can represent the genform $\Psi$ as the sum
$$
\Psi = \Psi e = \Psi(t+t^\prime)=\Psi t + \Psi t^\prime = \phi + \phi^\prime,
$$
where $\phi\in I(t)$, $\phi^\prime\in I(t^\prime)$. Let us substitute this decomposition of $\Psi$ into Cauchy problem (\ref{mod:Dirac:eq2}),(\ref{init:cond2}). We see that the genforms $\phi,\phi^\prime$ satisfy two independent Cauchy problems. Namely, the genform $\phi\in I(t)$ satisfies Cauchy problem (\ref{mod:Dirac:eq1}),(\ref{init:cond1}) and the genform $\phi^\prime\in I(t^\prime)$ satisfies the Cauchy problem
\begin{eqnarray}
&& h^\mu\partial_\mu\phi^\prime + i m \phi^\prime = 0,\quad x^1>0\label{mod:Dirac:eq3}\\
&& \phi^\prime = 0,\quad x^1=0.\label{init:cond3}
\end{eqnarray}
Let us multiply the left hand part of system of equations (\ref{mod:Dirac:eq3}) from left by $\beta = e^1$. Then we get the system of equations that can be written in the form of Friedrichs symmetric hyperbolic system of equations of first order. A Cauchy  problem for a Friedrichs symmetric hyperbolic system of equations is well posed in Hadamar sense. Hence, a solution of Cauchy problem (\ref{mod:Dirac:eq3}),(\ref{init:cond3}) exists and is unique.
So $\phi^\prime \equiv 0$ is the unique solution of (\ref{mod:Dirac:eq3}),(\ref{init:cond3}). This completes the proof of the theorem.


\begin{thebibliography}{99}
\bibitem{tr_mian2019}Marchuk N.G.: A generalization of Yang-Mills equations, Proc. Steklov Inst. Math., 306 (2019)


\bibitem{TMPh2019} Marchuk N.G.: One class of relativistic invariant systems of equations of first order, Theoret. and Math. Phys., (2020, in press)


\bibitem{Hestenes} Hestenes D.: Space-Time Algebra, Gordon and Breach, New York, (1966)


\bibitem{ILeq} Iwanenko D., Landau L.: Zur theorie des magnetischen electrons. I, Zeitschrift fur
Physik, Bd.48, 340-348, (1928)

\bibitem{Kahler}Kahler E.: Rendiconti di Mat. (Roma) ser. V, 21, 425 (1962)


 \bibitem{Atiyah}  Atiyah M.: Vector Fields on Manifolds,
Arbeitsgemeinschaft f\"ur Forschung des Landes
Nordrhein-Westfalen, Heft 200, (1970)


 \bibitem{BennTucker} Benn I.M., Tucker R.W.:  An introduction to spinors and geometry with applications to physics, Bristol, (1987)

\bibitem{ObSol}  Obukhov Yu. N.,  Solodukhin S. N.:
Reduction of the Dirac equation and its connection with the Ivanenko-Landau-Kahler equation, Theoret. and Math. Phys.,  94:2, 198-210 (1993)




\bibitem{Mybook2018} Marchuk N.: Field theory equations, Amazon, CreateSpace, (2012)

\bibitem{Godunov1979}Godunov S.K.: Uravneniya matematicheskoy fiziki (Equations of mathematical physics), Moscow, Nauka (1979)


\bibitem{Fri} Friedrichs K.O.\: Symmetric hyperbolic linear differential equations, Comm. pure and Appl. Math., v.7, 2, 345-391. (1954)


\bibitem{Miz} Mizohata S.:  The Theory of Partial Differential Equations, Cambridge University Press, (1979)
\end{thebibliography}
\end{document}